# Optical absorption and emission mechanisms of single defects in hexagonal boron nitride

Nicholas R. Jungwirth and Gregory D. Fuchs

Cornell University, Ithaca, New York 14853, USA

**Abstract:** We investigate the polarization selection rules of sharp zero-phonon lines (ZPLs) from isolated defects in hexagonal boron nitride (h-BN) and compare our findings with the predictions of a configuration coordinate model involving two electronic states.  Our survey, which spans the spectral range ~550-740 nm, reveals that, in disagreement with a two-level model, the absorption and emission dipoles are often misaligned.  We relate the dipole misalignment angle ($\Delta\theta$) to the ZPL Stokes shift ($\Delta E$) and find that $\Delta\theta \approx 0°$ when $\Delta E$ corresponds to an allowed h-BN phonon frequency and that $0° \leq \Delta\theta \leq 90°$ when $\Delta E$ exceeds the maximum allowed h-BN phonon frequency.  Consequently, a two-level configuration coordinate model succeeds at describing excitations mediated by the creation of one optical phonon but fails at describing excitations that require the creation of multiple phonons.  We propose that direct excitations requiring the creation of multiple phonons are inefficient due to the low Huang-Rhys factors in h-BN and that these ZPLs are instead excited indirectly via an intermediate electronic state.  This hypothesis is corroborated by polarization measurements of an individual ZPL excited with two distinct wavelengths that indicate a single ZPL may be excited by multiple mechanisms.  These findings provide new insight on the nature of the optical cycle of novel defect-based single photon sources in h-BN.

Wide bandgap semiconductors host point defects, or color centers, that can feature optical and spin properties that are useful for applications in quantum optics, precision sensing, and quantum information technology [1–5]. Some color centers, such as the nitrogen vacancy (NV) center in diamond [6–11], are bright enough to be investigated in the single defect limit using single-molecule microscopy techniques [12,13]. While diamond is the most celebrated host material, the last several years have witnessed the discovery of defect-based single photon sources in SiC [1,14–20], ZnO [21–26], GaN [27], WSe$_2$ [28–30], WS$_2$ [31], and hexagonal boron nitride (h-BN) [32–45]. The latter three materials exist as two-dimensional monolayers and layered solids, thus offering the possibility of integrating single-photon sources with van der Waals heterostructure devices for tuning and other control. Defect emission in h-BN can be ultrabright [32], have a narrow linewidth [33], be tuned [39], and remain photostable up to 800 K [41]. These positive attributes have sparked strong interest in h-BN defects from research groups around the world [32–45]. Despite this surge of interest, most works have focused on characterizing the phenomenology of h-BN emission, leaving open several difficult to answer questions regarding the fundamental nature of h-BN quantum emitters. These include the structural origin of the defect(s) responsible for single photon emission, the reason(s) for the broad distribution of zero-phonon line (ZPL) energies ($E_{ZPL}$), the spin properties, and the physical mechanism(s) involved in the defect's optical cycle.

In this work we address the transition mechanism(s) involved in the defect's optical cycle. We perform spectrally resolved polarization measurements of optical absorption and emission at cryogenic temperatures and compare our findings with the predictions of a configuration coordinate model. We find that when the Stokes shift of the ZPL is less than the maximum phonon energy in h-BN, the defect's polarization properties are well-explained by a configuration coordinate model with two electronic states. Conversely, when the Stokes shift of the ZPL exceeds the maximum phonon energy, a configuration coordinate model with two electronic states fails at explaining the observed behavior. These findings suggest that ZPL emission may be mediated by an intermediate electronic state. This explanation is supported by polarization measurements performed with lasers of different energies that verify a single ZPL may be excited via multiple mechanisms. Our findings, which provide new insight on the optical properties and level structure of h-BN defects, are key for designing future experiments and applications.

The fundamental mechanism governing non-resonant absorption and emission from point defects has been known for some time [46] and is illustrated in the configuration coordinate diagram shown in Fig. 1. In this model, a defect may undergo incoherent transitions to and from an electronic ground state ($\mu$) and an electronic excited state ($\mu^*$) that are mediated by lattice phonons. Note that although only one phonon frequency $\omega, \omega^*$ is depicted for each electronic state $\mu, \mu^*$, this model is readily adapted to include additional phonon modes [47,48]. In the diagram, the horizontal axis corresponds to the nuclear coordinate $Q$ that specifies the lattice configuration and the vertical axis corresponds to the total energy of the defect-lattice system. The zero-phonon lattice configuration in the excited state, $Q_0^* = \langle \mu^*|Q|\mu^*\rangle$, differs from that of the ground state, $Q_0 = \langle \mu|Q|\mu\rangle$, because each state produces a unique electrostatic potential. Each optical cycle begins with the system in the electronic state $\mu$ and a vibronic state occupation of $n$ phonons, with a probability governed by the Bose-Einstein distribution. Following the absorption of an optical photon, the system rapidly thermalizes and the excited state $\mu^*$ with $n^*$ phonons is occupied. In the Frank-Condon approximation, where the fast electronic rearrangement precedes the slower lattice relaxation, the transition rate from to $(\mu^*, n^*)$ to $(\mu, n)$ is proportional to

$$|\langle\mu|r|\mu^*\rangle|^2 \left|\int dQ \phi_{\mu,n}^*(Q)\phi_{\mu^*,n^*}(Q)\right|^2, \tag{1}$$

where $\phi_{\alpha,m}(Q)$ is the $m$-phonon lattice wave function when the defect is in electronic state $\alpha$. Emission into the ZPL corresponds with $n = n^*$ transitions, where no phonons are created or annihilated. All other transitions contribute to the phonon sideband.

In Equation 1 the first term is the dipole matrix element of the transition. This term determines the polarization selection rules for absorption and emission and is symmetric under time reversal. Consequently, this model predicts *identical* polarization properties for absorption and emission. Additionally, because the symmetry properties of $\mu$ and $\mu^*$ are determined by the defect's crystallographic point group, the transition dipoles are typically aligned parallel or perpendicular to distinct crystallographic directions. The second term in Equation 1, the Frank-Condon factor $F_n^{n^*}$, is the overlap integral between displaced harmonic oscillators. This term determines the lineshape of the absorption and emission bands. For linear modes $\omega = \omega^*$ and the Frank-Condon factor becomes

$$F_n^{n^*} = e^{-S} S^{n-n^*} \left(\frac{n^*!}{n!}\right) \left(L_{n^*}^{n-n^*}(S)\right)^2, \qquad (2)$$

where $L_{n^*}^{n-n^*}$ are the associated Laguerre polynomials and $S$ is a measure of defect-lattice coupling called the Huang-Rhys factor. In natural units $S = \frac{1}{2} m_{eff} \omega (Q_0 - Q_0^*)^2$, where $m_{eff}$ is the effective mass of the mode. At temperature $T = 0$, $F_n^0 = \frac{e^{-S} S^n}{n!}$ and the number of phonons created in a radiative transition is Poisson distributed about an average value of $S$. In this limit the relative spectral weight of the ZPL is $e^{-S}$, which is often termed the Debye-Waller factor. Because $F_n^{n^*}$ is symmetric under time reversal, the absorption and emission bands of a transition are mirror reflections of one another about $E_{ZPL}$.

To test the success of the Huang-Rhys model at describing the optical properties of isolated h-BN defects we performed polarization spectroscopy using a house-built confocal microscope [25] (see Supporting Information for microscope and sample details). Fig. 2a is a $T = 5K$ emission spectrum of defect that reveals the presence of a narrow ZPL at ~584 nm. The two-photon correlation function, $g^{(2)}(\tau)$, of the collected photons is shown as an inset. For this measurement collection was limited to the spectral region of Fig. 2a using optical filters. The antibunching dip at $\tau = 0$ extends well below 0.5, verifying that the ZPL corresponds to single photon emission from a single defect transition. To investigate the polarization properties of absorption, we rotate the polarization of the exciting light and monitor the total fluorescence intensity. The result of this absorption measurement is shown as the green triangles in Fig. 2b. Fixing the polarization of the exciting light to maximize the fluorescence, we determine the polarization of the emitted photons using a polarization analyzer placed in the collection path of the microscope. The result of this emission measurement is shown as the red circles in Fig. 2b. The solid lines are best fits to the data using the equation

$$A + B \cos^2\left[\frac{\pi}{180}(\theta - \langle\theta\rangle)\right], \tag{3}$$

where $\langle\theta\rangle$ in each fit is the orientation of the absorption or emission dipole spectrally averaged over the collection window. As predicted by Equation 1, we find that the maxima of absorption and emission are aligned for this defect. Additionally, we have shown previously that the temperature dependence of the ZPL intensity in h-BN is well-modeled by the temperature-dependent Debye-Waller factor [33]. Thus, we conclude that the configuration coordinate model is a good description of the observed properties for the defect shown in Fig. 2.

A survey of defect ZPLs that span an appreciable energy range reveals that, in contrast to the data shown in Fig. 2, the absorption dipole is frequently not aligned parallel to the emission dipole. To verify that the misalignment we observe is not an experimental artifact resulting from the wavelength- and polarization-dependent retardances introduced by optical elements in the microscope, we performed spectrally resolved polarization measurements (See Supporting Information). For the absorption measurement we vary the polarization of the exciting light and record an emission spectrum at each angle. This measurement produces $I_{abs}(\theta, E)$, where $I_{abs}(\theta, E)\delta E$ is the number of photons collected in the energy range $(E, E + \delta E)$ when the exciting light is polarized at $\theta$. For each energy $E'$ we fit $I_{abs}(\theta, E')$ to Equation 3 to obtain $\theta_{abs}(E')$, which is the polarization angle that maximizes $I_{abs}(\theta, E)$ when $E = E'$. For the emission measurement we fix the polarization angle of the exciting light to $\theta_{abs}(E_{ZPL})$ and record an emission spectrum for a series of positions of the polarization analyzer in the collection path. In an analogous fashion to the absorption case we obtain $I_{emit}(\theta, E)$ and $\theta_{emit}(E)$ for the emitted light. For the case of emission we apply a calibration to $\theta_{emit}(E)$ to correct for wavelength- and polarization-dependent retardances (see Supporting Information) introduced by the collection path of the confocal microscope.

Fig. 3a is a 2D image plot of $I_{emit}(\theta, E)$ for a single defect with a ZPL at 2.06 eV (603 nm) that is excited by 2.33 eV (532 nm) light. The red trace in Fig. 3b is the unpolarized emission spectrum, $I_{emit}(E)$, obtained by vertically summing the columns in the 2D image. The one- and two-optical-phonon sidebands are evident at ~1.88 eV and ~1.7 eV, respectively, corresponding to a phonon energy of ~180 meV. The red circles in Fig. 3c are the spectrally averaged polarization of the emitted light, $I_{emit}(\theta)$, obtained by horizontally summing the rows in the 2D image. Lastly, the red trace in Fig. 3a corresponds to the calibrated $\theta_{emit}(E)$ and indicates that, consistent with Equation 1, the polarization state of photons emitted into the ZPL is the same as for those emitted into the phonon sideband. We also measured $I_{abs}(\theta, E)$ (data not shown) and have included $\theta_{abs}(E)$ as the green trace in Fig. 3a. This trace indicates that the ZPL and phonon sideband intensities are maximized by the same polarization angle of the exciting light. However, in disagreement with Equation 1, the absorption and emission dipoles are not aligned (e.g. $\Delta\theta = |\theta_{emit}(E_{ZPL}) - \theta_{abs}(E_{ZPL})| \neq 0$), suggesting that additional processes may be involved in this defect's optical cycle.

To better understand the failure of the model we measured $\Delta\theta$ for 103 ZPLs distributed across the region 550-740 nm. Fig. 4a is a scatter plot relating the dipole misalignment $\Delta\theta$ of a ZPL to its Stokes shift, defined as $\Delta E = E_{exc} - E_{ZPL}$, where $E_{exc}$ is the energy of the exciting light. Incidentally, when $\Delta E$ is less than ~200 meV the data points are clustered near small values of $\Delta\theta$, as predicted by the configuration coordinate model. Conversely, when $\Delta E$ exceeds ~200 meV the data points are broadly distributed between 0° and 90°. Therefore ~200 meV corresponds to a critical Stokes shift above which Equation 1 often fails. We will now frame this critical energy in terms of $F_n^{n^*}$ and the h-BN bulk phonon density of stares (DOS).

At cryogenic temperatures the absorption band $W(E)$ resulting from a single phonon mode is related to the Frank-Condon factor by the expression $W(E) \approx W_0 \sum_{n^*} F_0^{n^*} f(E, n^*)$, where $W_0$ is the oscillator strength and $f(E, n^*)$ is the lineshape of the $n^*$-phonon sideband. In Fig. 4b we plot the theoretical $W(\Delta E)$ for two of the defects investigated. To determine $W(\Delta E)$ we first converted the experimentally measured luminescence spectrum $I_{emit}(E)$ to its associated emission band by the conversion factor $E^{-3}$ that accounts for the energy-dependent density of optical states [49]. Assuming linear phonon modes, we obtain $W(E)$ by reflecting the emission band about $E_{ZPL}$. To enable direct comparison with Fig. 4a, we finally shift $W(E)$ by $-E_{ZPL}$ to obtain $W(\Delta E)$. This comparison is meaningful because, for defects whose absorption is described by Equation 1, $W(\Delta E)$ approximates how strongly a ZPL with a particular Stokes shift will couple to the exciting light. Evidently, the regions of strongest absorption correspond to Stokes shifts of ~160-200 meV. Fig. 4a indicates that defects with a ZPL in this spectral range are indeed explained by the configuration coordinate model.

Here we compare the energies just identified to the relevant phonon energies in h-BN [50]. In this comparison we only consider bulk lattice modes and neglect truly local phonon modes. This focus is justified post hoc by the success of appealing to these modes. The lowest energy modes are acoustic phonons, and we have shown previously that in-plane acoustic phonons exponentially broaden defect emission in the vicinity of the ZPL as temperature is increased [33]. Consequently, acoustic phonons are relevant for the optical properties of defects in h-BN. However, acoustic phonons in h-BN range in energy from ~0-107 meV, and are likely not the dominant mode responsible for the absorption band peaks evident in Fig. 4b. Optical phonon energies, on the other hand, extend from ~72-203 meV and are therefore strong candidates for phonon-mediated absorption and emission. Specifically, out-of-plane optical phonons range in energy from ~72-145 meV whereas in-plane optical phonons range from ~150-203 meV. Only the energies of in-plane optical phonons match the energies identified earlier in Fig. 4a and b. To aid in visualizing these energies we have highlighted three regions labeled I, II, and III in Fig. 4a,b that correspond to the absorption band of one, two, and three in-plane optical phonons, respectively. Note that only in region I is the configuration coordinate model successful, consistent with the low Huang-Rhys factors reported previously [33,42].

Here we propose an absorption and emission mechanism to explain the broad $\Delta\theta$ distribution that incorporates, rather than contradicts, the configuration coordinate model presented earlier. In Fig. 1 direct absorption between two electronic states is mediated by lattice phonons. This scenario of direct absorption is again depicted on the left of Fig. 4c, where the vibronic states of the lattice are represented as a blurred continuum. Alternatively, the two electronic states that produce a ZPL may be coupled via a third intermediate electronic state. Note that the intermediate electronic state may be intrinsic to the defect or may originate from a neighboring defect. This case of indirect absorption is depicted on the right of Fig. 4c. Here transitions between any pair of electronic states are still described by the configuration coordinate model. However, because the electronic states coupled by the exciting light differ from those that produce the ZPL, we no longer anticipate $\Delta\theta = 0°$.

Although the indirect absorption mechanism correctly predicts a broader $\Delta\theta$ distribution, it does not predict the shape of the distribution, shown in Fig. 4d.

Specifically, if the electronic states are all crystallographically related, group theoretic considerations [25,51] predict that $\Delta\theta = 0°$ for direct absorption and $\Delta\theta \epsilon \{0°, 30°, 60°, 90°\}$ for indirect absorption. However, the results in Fig. 4d do not reveal clustering at these values but rather suggest a flat distribution with clustering at 0°. We propose two explanations for this disagreement. Firstly, we note that our measurement of $\theta_{abs}(E_{ZPL})$ and $\theta_{emit}(E_{ZPL})$ is sensitive only to the projection of the absorption and emission dipole into the plane perpendicular to the axis of the exciting light. Consequently, because the h-BN flakes we investigated are often tilted relative to the substrate (see Supporting Information), the $\Delta\theta$ we measure can differ from the true dipole misalignment. Secondly, it is possible that direct and indirect absorption mechanisms may coexist for a particular ZPL. In this scenario $\theta_{emit}(E_{ZPL})$ is related to the true emission dipole orientation and $\theta_{abs}(E_{ZPL})$ is actually a weighted average over all absorption mechanisms. To test whether a particular ZPL may originate from two distinct mechanisms we acquired spectrally-resolved polarization measurements using both 532 nm and 473 nm light for excitation. Fig. 5a is a magnified view of a ZPL at ~577 nm excited using 532 nm (green trace) and 473 nm (blue trace) light, corresponding to Stokes shifts of ~182 meV and ~472 meV, respectively. The two spectra overlap well, verifying that each wavelength may excite the same ZPL. In Fig. 5b we plot $\theta_{emit}(E)$ (red trace), $\theta_{abs}(E)$ for 532 nm excitation (green trace), and $\theta_{abs}(E)$ for 473 nm excitation (blue trace). Incidentally, $\Delta\theta \approx 0°$ for 532 nm excitation and $\Delta\theta \approx 50°$ for 473 nm excitation. For this defect, the 532 nm excitation is well described by direct absorption whereas the 473 nm excitation is explained by indirect absorption. Therefore, a particular ZPL may indeed be excited via multiple mechanisms.

In conclusion, we made polarization measurements of absorption and emission for 103 isolated defects in h-BN with ZPLs in the range ~550-740 nm. In contrast to the predictions of a Huang-Rhys model involving two electronic states, our survey reveals that the absorption and emission dipoles are frequently misaligned. We frame the dipole misalignment $\Delta\theta$ in the context of the Stokes shift of a ZPL ($\Delta E$), rather than its energy, and demonstrate that $\Delta E$ is a strong indicator of the likelihood that the absorption and emission dipoles will be parallel. In particular, if $\Delta E$ coincides with an allowed phonon energy in h-BN then $\Delta\theta \approx 0°$. Therefore direct absorption mediated by the creation a single phonon is efficient and is well described by the Huang-Rhys model with two electronic states. Alternatively, if $\Delta E$ exceeds the maximum phonon energy in h-BN then $0° \leq \Delta\theta \leq 90°$. We propose a mechanism to explain these observations whereby a defect may be excited indirectly through a third intermediate electronic state. This mechanism is supported by polarization measurements of a defect ZPL acquired using 532 nm and 473 nm excitation, which reveal that single defect's ZPL emission may be excited via multiple mechanisms. These comprehensive results form a key advance in our understanding of defect absorption and emission mechanisms in h-BN single defects.

**Supporting Information:** Experimental apparatus and spectrally-averaged polarization measurements; Spectrally-resolved polarization measurements; Sample details; Lifetime distribution; Visibility distribution

**Acknowledgements:**

We thank Brian Calderon for acquiring SEM images of annealed h-BN flakes. This work was supported by the National Science Foundation (DMR-1254530). We acknowledge use of the Cornell NanoScale Facility, a member of the National Nanotechnology Coordinated Infrastructure (NNCI), which is supported by the National Science Foundation (Grant ECCS-15420819). Additionally, we acknowledge the Cornell Center for Materials Research Shared Facilities, which are supported through the NSF MRSEC program (DMR-1120296).

Note: The authors declare no competing financial interest.

**FIG. 1:**

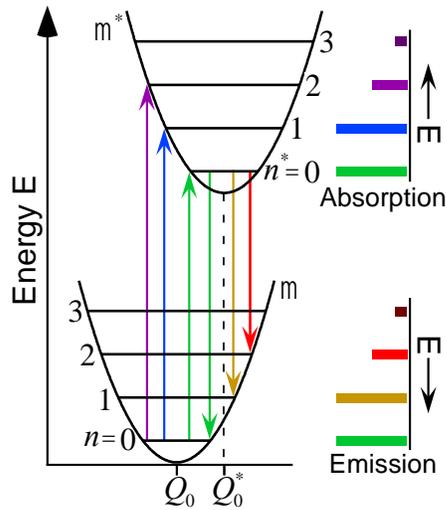

Configuration coordinate diagram with $S = 1$ illustrating phonon-mediated transitions to and from a defect's electronic ground state $\mu$ and electronic excited state $\mu^*$ at zero temperature. While in electronic state $\mu$ with $n = 0$ phonons, the defect may absorb an optical photon and enter state $(\mu^*, n^*)$. Following rapid thermalization to the vibronic ground state the defect may radiatively relax from state $(\mu^*, 0)$ to $(\mu, n)$ with a probability given by the Frank-Condon factor. For linear modes $(\omega = \omega^*)$ the absorption and emission bands, shown on the right, are mirror reflections of one another.

**FIG 2:**

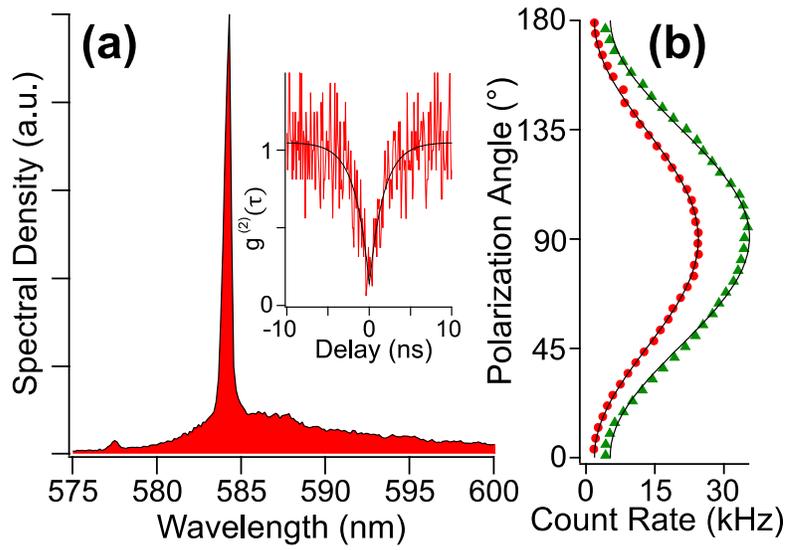

(a) Emission spectrum of a defect revealing a sharp ZPL at ~584 nm. This ZPL corresponds to single photon emission, as verified by the antibunching dip in $g^{(2)}(\tau)$ shown as an inset. (b) The polarization profiles for absorption (green triangles) and emission (red circles) are aligned.

**FIG 3:**

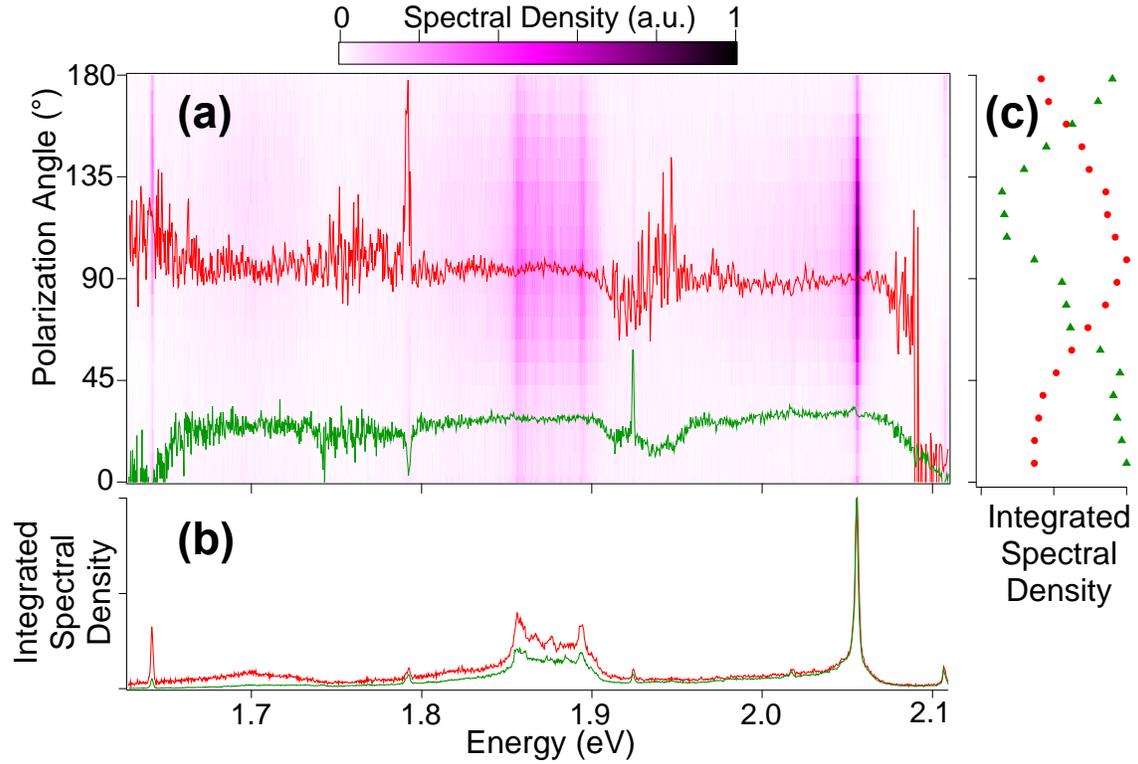

(a) 2D Image plot of $I_{emit}(\theta, E)$ with a ZPL at ~2.06 eV. The average polarization of photons emitted with energy $E$, $\theta_{emit}(E)$, is shown as the red trace. Note that photons emitted into the ZPL have the same polarization as those emitted into the one- and two-phonon sidebands at 1.88 eV and 1.7 eV. The unpolarized emission spectrum $I_{emit}(E)$ and the spectrally average polarization profile $I_{emit}(\theta)$, obtained by integrating the columns and rows of (a), are included as the red data in (b) and (c), respectively. The green data in (a), (b), and (c) are the analogous measurements for absorption, obtained from $I_{abs}(\theta, E)$ (not shown). In contrast to Fig. 2, $\theta_{emit}(E_{ZPL}) \neq \theta_{abs}(E_{ZPL})$.

**FIG 4:**

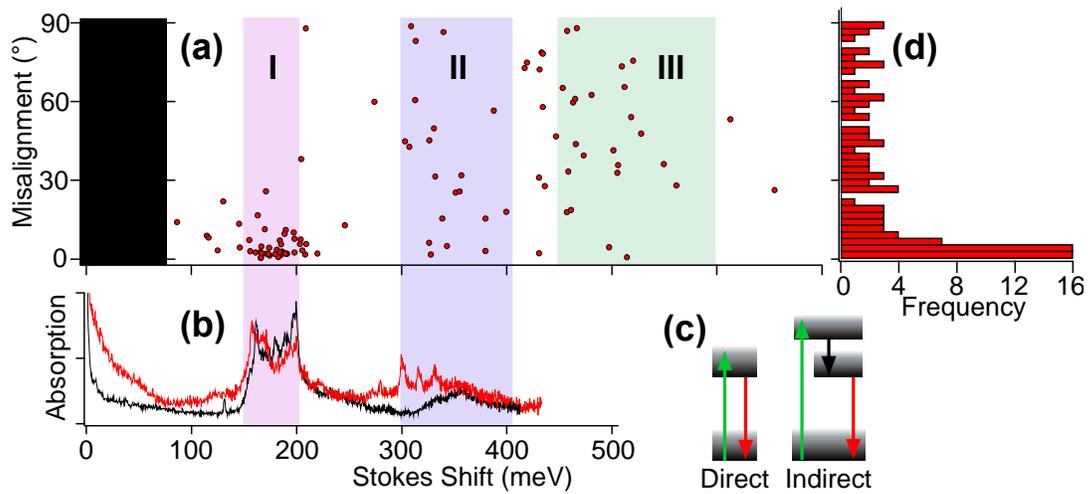

(a) Scatter plot relating the misalignment angle between the absorption and emission dipole of a ZPL ($\Delta\theta$) to its Stokes shift from the exciting light ($\Delta E$) for 103 defects. The black box that extends to ~75 meV represents ZPL energies that could not be studied due to our selection of optical filters. The shaded region labeled "I" corresponds to the energies of in-plane optical phonons and points in this region are clustered near $\Delta\theta = 0$. ZPLs in Region II and III may be excited via the creation of two and three optical phonons, respectively. Only points in Region I and below agree with a configuration coordinate model involving two states. (b) The theoretical absorption band, $W(\Delta E)$, of two defects reveals peaks in Region I and II, verifying that in-plane optical phonons are relevant for absorption and emission. (c) Two energy-level diagrams illustrating direct (left) and indirect (right) excitation mechanisms. The left diagram is equivalent to Fig. 1 and predicts $\Delta\theta = 0$ whereas the right diagram allows for a broad $\Delta\theta$ distribution. (d) Histogram of all $\Delta\theta$ values shown in (a).

**FIG 5:**

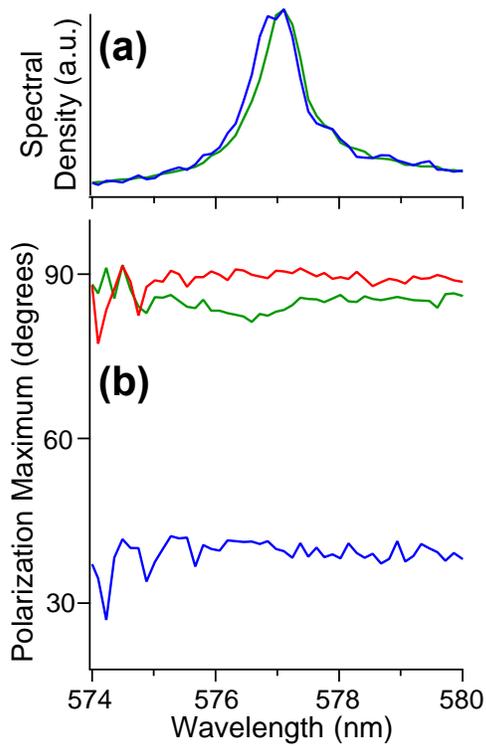

(a) Magnified emission spectrum of a ZPL at ~577 nm excited with 473 nm (blue trace) and 532 nm (green trace) light indicating that both energies may excite the transition. (b) Spectrally resolved polarization measurements of $\theta_{abs}(E)$ for excitation with 532 nm light (green trace) and $\theta_{emit}(E)$ (red trace) reveal that $\Delta\theta(E_{ZPL}) \approx 0$. The analogous measurement using 473 nm excitation (blue trace) indicates a large misalignment between the 473 nm absorption dipole and the emission dipole. Therefore an individual ZPL may be excited via multiple mechanisms.

# Optical Absorption and Emission Mechanisms of Single Defects in Hexagonal Boron Nitride: Supplementary Information

Nicholas R. Jungwirth and Gregory D. Fuchs

Cornell University, Ithaca, New York 14853, USA

# I. EXPERIMENTAL APPARATUS AND SPECTRALLY-AVERAGED POLARIZATION MEASUREMENTS

Fig. S1 is a schematic of the house-built confocal microscope used in this work to study point defects in hexagonal boron nitride (h-BN). For steady-state measurements we use either a continuous wave (CW) 532 nm laser or a CW 473 nm laser for excitation. For lifetime measurements (shown in **Section IV** below) we use a pulsed 532 nm laser with an 80 kHz repetition rate and a 350 ps pulse width. To create an arbitrary linear polarization state of the exciting light we use a fixed linear polarizer (FP1) followed by a rotatable half wave plate (HWP1). Between HWP1 and the 0.7 NA microscope objective (MO) the exciting light inherits a polarization-dependent retardance from elements in the optical path that both rotates the polarization state and reduces the extinction ratio. We compensate for these effects using a fixed wave plate (FWP1) selected to introduce an appropriate correcting retardance at 532 nm. To calibrate the polarization state of the excitation path we place a polarization analyzer in front of the MO and measure the polarization state of the exciting light for a series of positions of HWP1.

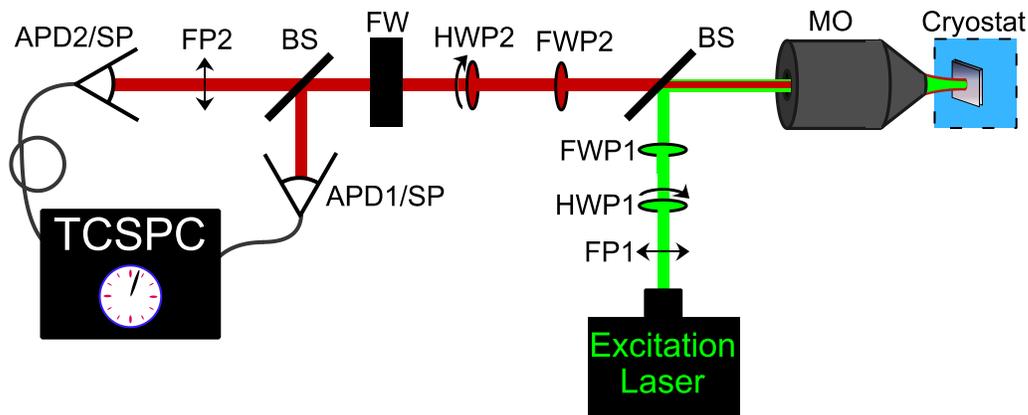

**Fig. S1:** Schematic of confocal microscope used in this work.

Optically active defects may absorb the exciting light and emit photons that are collected by the MO. We use a beam splitter (BS), two single photon detectors (APDs), and a time correlated single photon counting module (TCSPC) to measure the two-photon correlation function, $g^{(2)}(\tau)$, of the collected photons. Photons directed towards APD1 are detected independent of their polarization state. Thus we probe the spectrally averaged absorption dipole, $\langle \theta_{abs} \rangle$, by rotating HWP1 and monitoring the count rate on APD1. To determine the average polarization state of the collected photons we repeat the procedure used for absorption. We first correct for the wavelength- and polarization-dependent retardances of the collection path using FWP2. Next the polarization state of the collected photons is rotated by achromatic HWP2. Finally, the collected photons may pass through FP2 with a probability determined by their polarization state. Thus we probe the spectrally averaged emission dipole, $\langle \theta_{emit} \rangle$, by rotating HWP2 and monitoring the count rate on APD2. To calibrate the average polarization state of the collection path we direct 633 nm light from APD2 towards the MO, place a polarization analyzer in front of the MO, and measure the polarization state of the light for a series of positions of HWP2. For all measurements a particular zero-phonon line (ZPL) may be spectrally isolated by selecting an appropriate combination of long- and short-pass filters from the filter wheel (FW).

## II. SPECTRALLY-RESOLVED POLARIZATION MEASUREMENTS

In addition to the conventional, spectrally-averaged polarization measurements discussed in **Section I**, we also performed spectrally-resolved polarization measurements for two primary reasons. First, because the polarization state of the collection path is wavelength dependent, and because the distribution of zero-phonon line energies ($E_{ZPL}$) is broad, a spectrally-resolved calibration is essential to ensure the polarization state we measure for a particular ZPL energy faithfully represents the true polarization state. Secondly, because multiple distinct ZPLs may be simultaneously excited within the same h-BN flake, spectrally-resolved measurements are essential to properly distinguish the polarization properties of each ZPL. For the absorption measurement we rotate HWP1 and, rather than detect the light at APD1, we direct the light to a spectrometer (SP). For each position of HWP1 we record an emission spectrum to directly obtain $\tilde{I}_{abs}(\phi, \lambda)$, where $\tilde{I}_{abs}(\phi, \lambda)\delta\lambda$ is the number of photons collected in the wavelength range $(\lambda, \lambda + \delta\lambda)$ when HWP1 is oriented at an angle $\phi$. We may convert $\tilde{I}_{abs}(\phi, \lambda)$ to $I_{abs}(\theta, E)$, where $I_{abs}(\theta, E)\delta E$ is the number of photons collected in the energy range $(E, E + \delta E)$ when the exciting light is polarized at angle $\theta$, via the conversion $I_{abs}(\theta, E) \propto \lambda^2 \tilde{I}_{abs}(\phi, \lambda)$. Note that for excitation measurements the calibration of **Section I** may be used to convert $\phi$ to $\theta$. At each fixed energy $E'$ we fit $I_{abs}(\theta, E')$ to the function

$$A + B \cos^2\left[\frac{\pi}{180}(\theta - \theta_{abs}(E'))\right], \tag{1}$$

where $\theta_{abs}(E')$ is the energy-dependent orientation of the absorption dipole.

Fig. S2a is an emission spectrum of a h-BN flake revealing the presence of two sharp ZPLs separated by ~20 nm. A spectrally-averaged polarization measurement would produce $\langle\theta_{abs}\rangle$, which represents the polarization state of the exciting light that maximizes the integrated fluorescence from both ZPLs. In contrast, a spectrally-resolved polarization measurement produces $\theta_{abs}(\lambda)$, which represents the polarization state of the exciting light that maximizes the spectral density at the wavelength $\lambda$. This measurement is shown in Fig. S2b and verifies that the absorption dipole of each ZPL is misaligned.

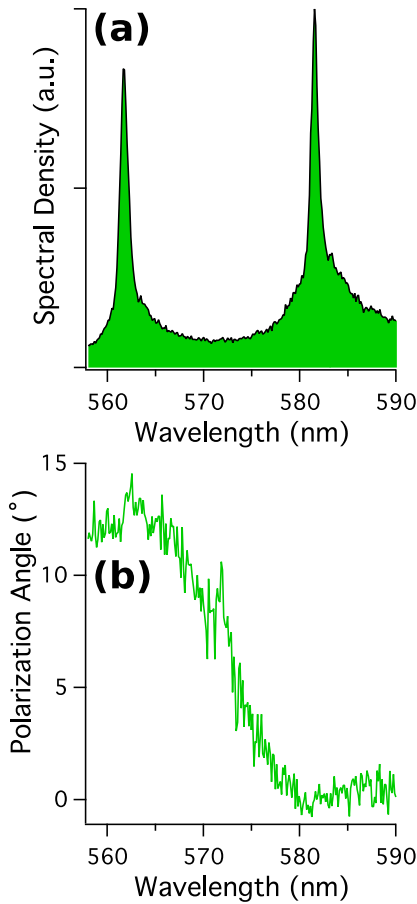

**Fig. S2:** (a) Emission spectrum of a h-BN flake revealing the presence of two distinct ZPLs. (b) Measurement of $\theta_{abs}(\lambda)$ for the flake that produced the spectrum in (a). Evidently distinct ZPLs with distinct polarization properties may be simultaneously excited.

After measuring $\theta_{abs}(E)$ for a particular ZPL we may fix the polarization state of the exciting light to $\theta_{abs}(E_{ZPL})$ and measure $\tilde{I}_{emit}(\phi, \lambda)$ by recording an emission spectrum at each orientation of HWP2. Analogous to the absorption case, $\tilde{I}_{emit}(\phi, \lambda)\delta\lambda$ represents the number of photons detected in the energy range $(\lambda, \lambda + \delta\lambda)$ when HWP2 is oriented at angle $\phi$. However, in contrast to the absorption case, because the retardances introduced by the collection path depend on both the wavelength and the polarization state of the emitted light, we can no longer use the spectrally-averaged calibration from **Section I** to convert $\phi$ to $\theta$. Consequently, to perform spectrally-resolved polarization measurements of emission we must apply a polarization- and wavelength-dependent calibration to $\phi$ to obtain $I_{emit}(\theta, E)$. To perform this calibration we place a broadband light source polarized at angle $\theta_{emit}$ at the objective and direct the light towards APD2. We then measure $\tilde{I}_{emit}(\phi, \lambda)$. In Fig. S3a we plot $2\phi_{emit}(\lambda) - \theta_{emit}$ when the light source is polarized at angles $\theta_{emit} \in \{0°, 30°, 60°, 90°, 120°, 150°\}$. Each trace represents the angular error one would obtain in a polarization measurement of photons with wavelength $\lambda$ that were emitted with a true polarization angle of $\theta_{emit}$ if a wavelength- and polarization-dependent calibration were not implemented. Note that the magnitude of this error is minimized by using a BS, rather than a dichroic mirror, to combing the excitation and collection paths. In Fig. S3b we plot the spectrally-resolved emission visibility $V_{emit}$ of the microscope at each polarization angle $\theta_{emit}$, where

$$V_{emit} = \frac{B}{B+2A}. \tag{2}$$

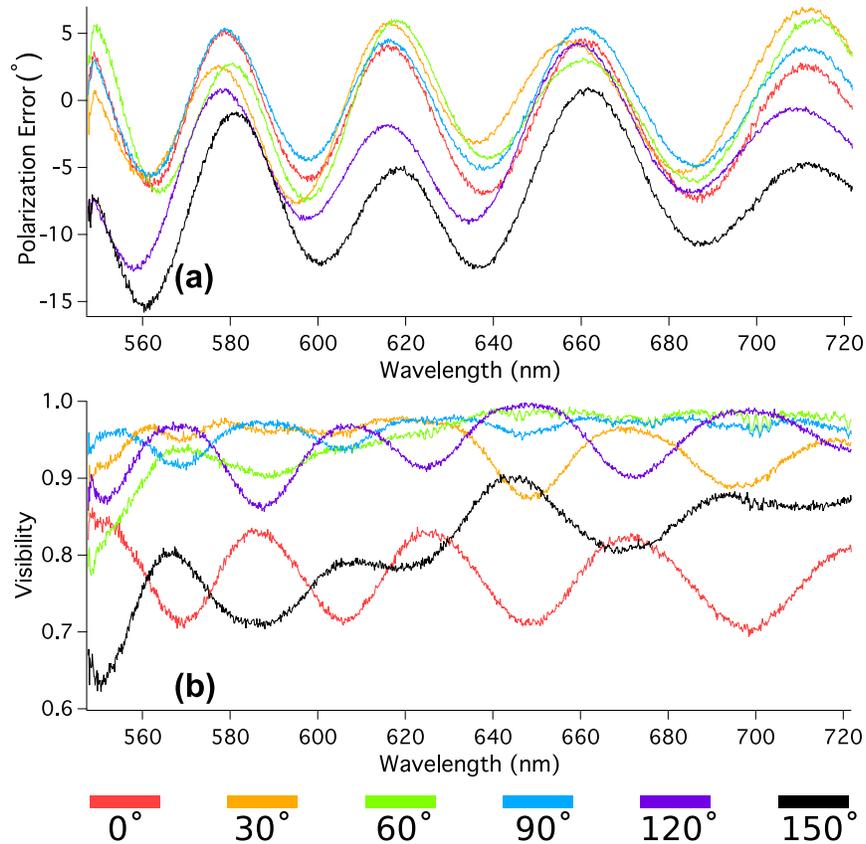

**Fig. S3:** (a) Calibration of the polarization state of the collected light that accounts for the wavelength-dependent properties of the collection path when the emitted light is initially polarized at angle $\theta_{emit}$. (b) Theoretical maximum measured visibility for light that is purely polarized at angle $\theta_{emit}$ when it exits the microscope objective. In both (a) and (b) traces corresponding to $\theta_{emit} \in \{0°, 30°, 60°, 90°, 120°, 150°\}$ are included.

## III. SAMPLE DETAILS

The h-BN flakes we investigated are commercially available from Graphene Supermarket. The as-received flakes are suspended in a 50/50 water/ethanol solution. We drop cast 25 $\mu L$ of solution onto a thermally oxidized silicon substrate and anneal the samples at 850°$C$ for 30 minutes under continuous nitrogen flow. We have previously characterized our samples via Raman spectroscopy and energy dispersive X-ray spectroscopy [1]. Fig S4a and b are two scanning electron microscopy (SEM) images of the as-prepared sample. Note that the flakes aggregate together and consequently many of the flakes are tilted with respect to the substrate.

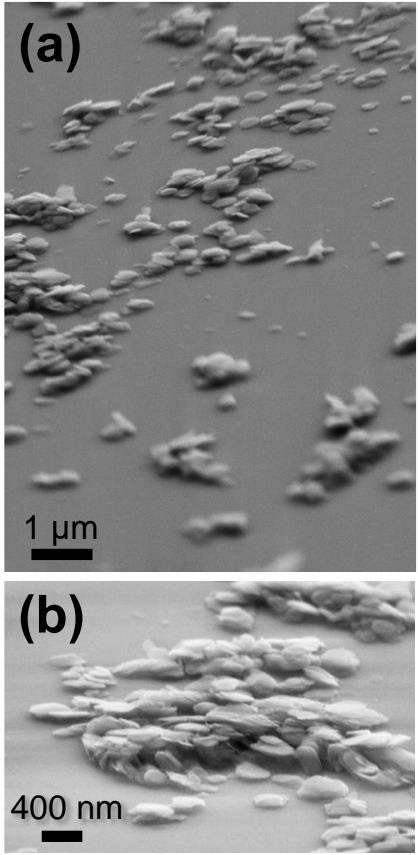

**Figure S4:** (a) SEM image of the prepared sample. (b) Magnified view of a cluster of h-BN flakes revealing multiple flake orientations.

The polarization measurements of $\theta_{abs}$ and $\theta_{emit}$ discussed in **Section II** are only sensitive to the projection of the absorption and emission dipole, respectively, into the plane of the substrate. Consequently, the measured dipole misalignment angle $\Delta\theta = |\theta_{abs} - \theta_{emit}|$ may differ from the true misalignment angle. This effect may be offset by investigating h-BN samples with known orientation.

## IV. LIFETIME DISTRIBUTION

For 36 of the investigated emitters we also measured the lifetime of the excited state using pulsed excitation. Fig. S5 is a scatter plot relating the ZPL wavelength to the excited state lifetime. Evidently most of the emitters investigated have a lifetime near 3 ns, but the excited state lifetime may be as low as 1.5 ns and as high as 8 ns.

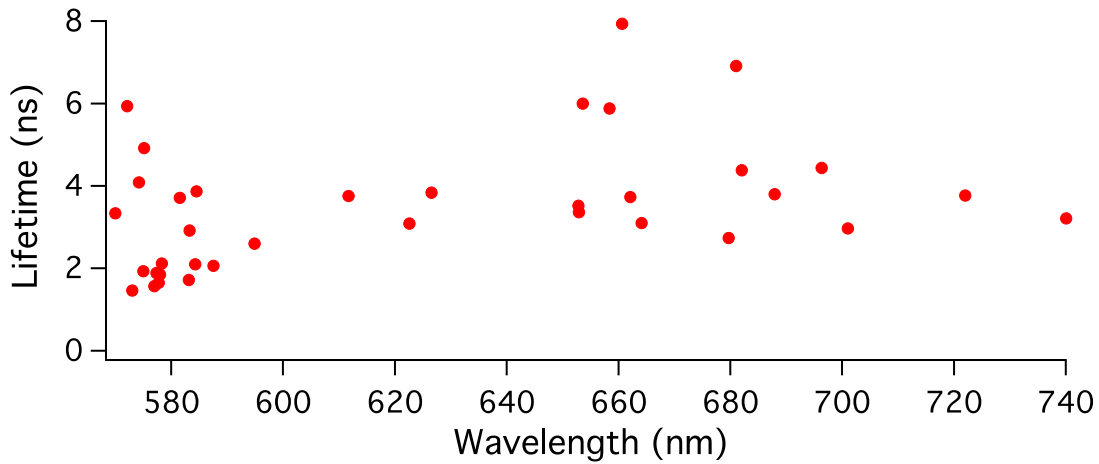

**Fig S5:** Scatter plot relating the ZPL wavelength to the excited state lifetime for 36 of the emitters investigated.

## V. VISIBILITY DISTRIBUTION

In the main text we provided evidence that a single ZPL may be excited both via direct absorption and indirect absorption with the efficiency of each mechanism being determined by the laser energy being used. We anticipate the visibility of absorption and emission to be similar for a direct transition [2]. For an indirect transition, however, we anticipate the emission visibility to exceed the absorption visibility. Fig. S6 is a scatter plot relating the absorption visibility and emission visibility for all the emitters investigated. Note that the emission visibilities have been corrected using the calibration shown in Fig. S3b to more closely approximate their true values. The solid line is a plot of the function $y = x$. Consequently any points above the line have an emission visibility that exceeds the absorption visibility. The majority of data points lie above the solid line, in support of the mechanism proposed in the main text.

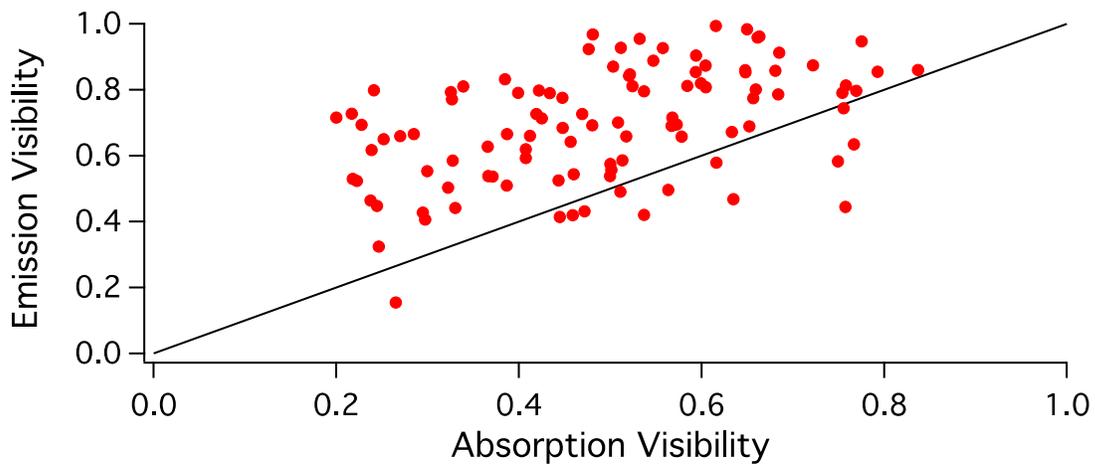

**Fig. S6:** Scatter plot relating the absorption and emission visibilities for all defects investigated.